\documentclass[%
 aip,
 amsmath,amssymb,
 reprint,%
]{revtex4-1}

\usepackage{graphicx}
\usepackage{dcolumn}
\usepackage{bm}

\usepackage[utf8]{inputenc}
\usepackage[T1]{fontenc}
\usepackage{mathptmx}
\usepackage{etoolbox}
\usepackage{longtable}

\makeatletter
\def\@email#1#2{%
 \endgroup
 \patchcmd{\titleblock@produce}
  {\frontmatter@RRAPformat}
  {\frontmatter@RRAPformat{\produce@RRAP{*#1\href{mailto:#2}{#2}}}\frontmatter@RRAPformat}
  {}{}
}%
\makeatother
\begin{document}

\preprint{AIP/123-QED}

%
%
%
%

\title[Design of a Doppler backscattering diagnostic for the Wisconsin HTS Axisymmetric Mirror (WHAM)]{Design of a Doppler backscattering diagnostic for the Wisconsin HTS Axisymmetric Mirror (WHAM)}

\author{E. Wikarta}
\email{Eduard\_Wikarta@a-star.edu.sg}
\affiliation{Future Energy Acceleration and Translation (FEAT) Centre, Agency for Science, Technology and Research (A*STAR), Singapore 138632, Singapore}

\author{U. Kumar}
\affiliation{Future Energy Acceleration and Translation (FEAT) Centre, Agency for Science, Technology and Research (A*STAR), Singapore 138632, Singapore}

\author{V.H. Hall-Chen}
\affiliation{Future Energy Acceleration and Translation (FEAT) Centre, Agency for Science, Technology and Research (A*STAR), Singapore 138632, Singapore}
\affiliation{School of Physical and Mathematical Sciences, Nanyang Technological University, Singapore 637371, Singapore}

\author{D. Endrizzi}
\affiliation{Realta Fusion, Madison, Wisconsin 53717, USA}

\author{S.J. Frank}
\affiliation{Realta Fusion, Madison, Wisconsin 53717, USA}

\author{C.M. Jacobson}
\affiliation{Realta Fusion, Madison, Wisconsin 53717, USA}

\author{X. Li}
\affiliation{Future Energy Acceleration and Translation (FEAT) Centre, Agency for Science, Technology and Research (A*STAR), Singapore 138632, Singapore}

\author{D.A. Sutherland}
\affiliation{Realta Fusion, Madison, Wisconsin 53717, USA}

\date{\today} 

\begin{abstract}
The Wisconsin HTS Axisymmetric Mirror (WHAM) is a compact high-field magnetic mirror. In such magnetic mirrors, cross-field transport is dominated by the flute instability (Endrizzi et al., 2023). To investigate density fluctuations associated with the flute instability, we designed a Doppler backscattering (DBS) diagnostic for WHAM, to be installed at the midplane port window. The diagnostic uses a two-channel tunable Ka-band (26.5--40 GHz) source and X-mode polarization. The azimuthal launch angle is set mechanically by rotating the external quasioptical assembly. As such, the system is reconfigurable during dedicated setup periods. 
Using the \textit{Scotty} beam-tracing code (Hall-Chen et al., 2022), we show that the proposed DBS system can measure density fluctuations with perpendicular wavenumbers $1 \leq k_\perp \leq 3~\mathrm{cm}^{-1}$ over radial locations $0.7 \leq \rho \leq 0.9$, where $\rho$ is the normalized radial coordinate. This is achieved with probe frequencies between 28 and 38.5 GHz, an elevation launch angle of $0^\circ$, and azimuthal launch angles in the range $1^\circ$--$3^\circ$. The selected configurations have low mismatch angle at cutoff, $|\theta_{m,c}|<1^\circ$.
The quasioptical system uses a Ka-band horn and a biconvex ultra-high molecular weight polyethylene lens, and satisfies the port-access constraints in WHAM. The planned microwave system has a monostatic, homodyne architecture based on two phase-coupled Ka-band microwave channels. These two channels will be for the transmitted signal and coherent local oscillator (LO) for IQ downconversion, respectively. 
As the two phase-coupled channels can be independently tuned or swept with a controlled frequency offset, the same microwave chain can also support profile-reflectometry measurements using cutoff-delay information.
\end{abstract}

\maketitle

\section{Introduction}
In axisymmetric linear mirrors, the dominant magnetohydrodynamic instability is the interchange mode, also known as the flute instability. \cite{Endrizzi:WHAM:2023} These modes can drive radial transport and degrade confinement, particularly near the plasma edge. Hence, measurements of the density fluctuations associated with these flute instabilities are important for validating stability and transport models in mirror plasmas. 

The Wisconsin HTS Axisymmetric Mirror (WHAM) is a compact, high-field axisymmetric mirror located at the University of Wisconsin--Madison. \cite{Endrizzi:WHAM:2023} WHAM is designed to test physics and technology pertaining to next-step linear fusion devices, including edge-biased limiters and end rings for generating sheared $E \times B$ flow to stabilize flute modes. To assess the effectiveness of this stabilization, a diagnostic capable of measuring flows and density fluctuations with good spatial and wavenumber resolution is needed. Such a diagnostic would also help to validate recent simulations, which predict that drift-cyclotron loss-cone activity in WHAM can lead to coherent edge-localized density perturbations and associated $E \times B$ flows, and may in turn contribute to cross-field transport. \cite{Post:mirror_instabilities:1966, Tran:DCLC:2025}

The Doppler backscattering (DBS) microwave diagnostic measures turbulent electron density fluctuations and perpendicular flows in fusion plasmas. \cite{Hirsch:DBS:2001, Hennequin:DBS:2004, Happel:DBS:2017, Schmitz:DBS:2018, Carralero:DBS:2021, Pratt:DBS:2022, Macwan:DBS:2024, Tong:DBS:2025, Shi:DBS:2025, Rienacker:DBS:2025, Liang:DBS:2026} DBS involves launching a microwave probe beam into the plasma. The beam is strongly refracted and subsequently backscattered primarily by density fluctuations near the cutoff location. \cite{Ruiz:DBS:2025, Conway:DBS:2025} The measured turbulence wavenumber is determined by the Bragg condition, $|k_\perp| \simeq 2K_c$, where $K_c$ is the probe-beam wavenumber at cutoff. \cite{Hall-Chen:beam:2022} Hence, the backscattered signal is proportional to the amplitude of turbulent density fluctuations of a particular wavenumber in the vicinity of the cutoff location. The Doppler shift of the backscattered signal in turn gives information about the plasma flow. \cite{Pratt:DBS:2022}

In this paper, we present a preliminary design of a DBS diagnostic for WHAM. The diagnostic is intended to measure density fluctuations near the plasma boundary using a single-channel tunable Ka-band source, X-mode polarization, and reconfigurable launch and receive geometry at the WHAM midplane port. The design must satisfy both plasma-physics requirements, including access to edge cutoff locations and the expected fluctuation wavenumber range, and engineering constraints imposed by the available port and quasioptical layout. The same hardware is also intended to be compatible with profile reflectometry operation.

The rest of this paper is organized as follows. Section~\ref{sec:equilibrium} describes the WHAM equilibrium and estimated cutoff locations. In Section~\ref{sec:measurement}, we show the expected measurement locations and wavenumbers, calculated using beam tracing. Finally, we present the quasioptical and hardware design in Section~\ref{sec:hardware}. 

\section{WHAM equilibrium} \label{sec:equilibrium} 
We consider a single representative WHAM scenario for the DBS design. The magnetic equilibrium is calculated with \textit{Pleiades}, a free-boundary Grad--Shafranov solver for axisymmetric mirror equilibria. \cite{Frank:WHAM:2026} The vacuum field is determined from the measured coil currents and machine geometry, while the plasma response is constrained by thomson scattering measurements of $n_e$ and $T_e$ and by diamagnetic flux loops measurements of the plasma's diagmagnetic response. This experimentally constrained equilibrium is used as the input to the \textit{Scotty} beam-tracing \cite{Hall-Chen:beam:2022} simulations.

\begin{figure}
    \includegraphics[width=0.47\textwidth]{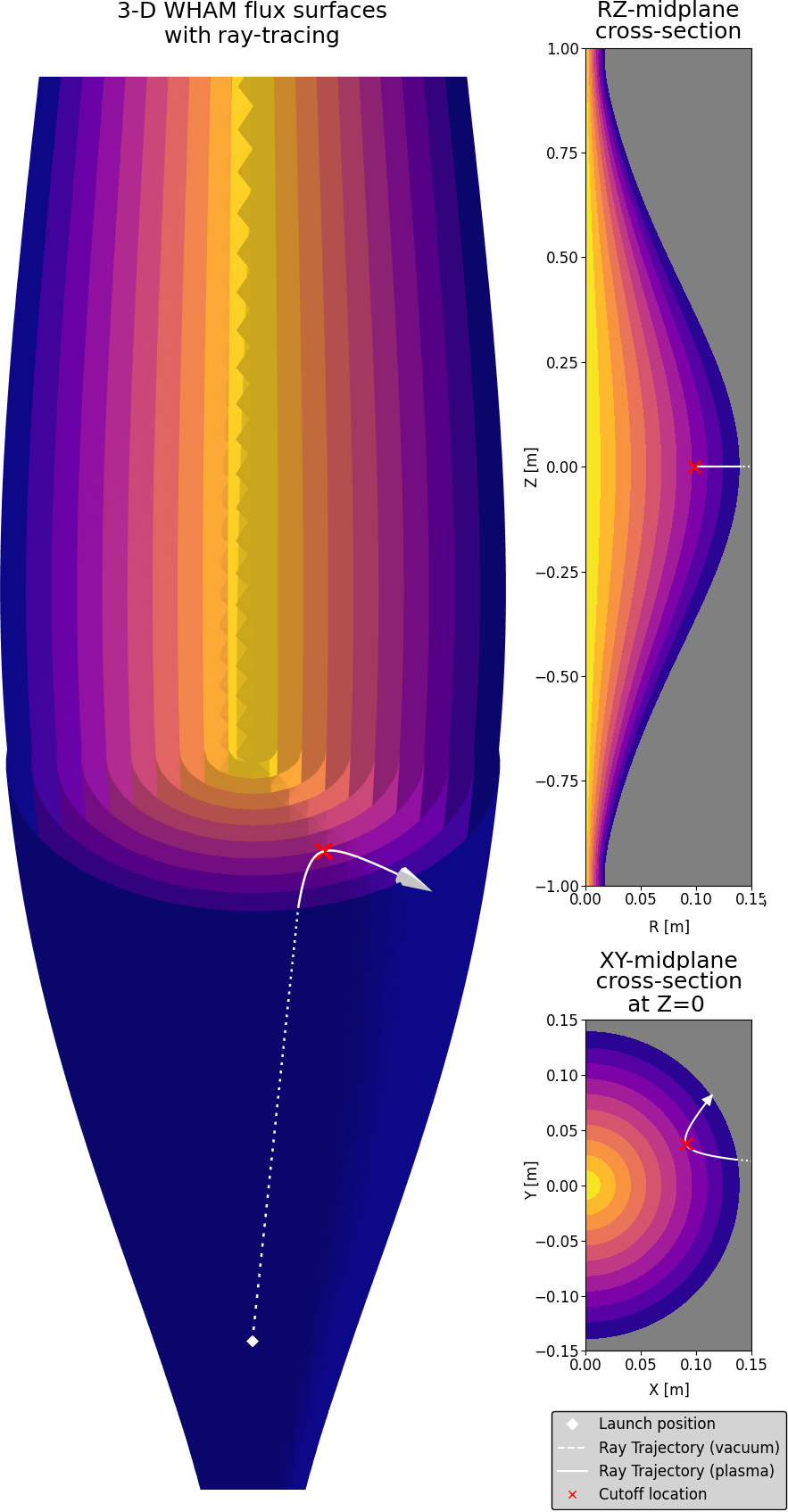}
    \caption{3-D plot (left) with the RZ-midplane (top right) and midplane, given by $Z=0$, (bottom right) cross-sections of the magnetic flux surfaces in WHAM. The dotted and solid white lines represent the ray trajectory in the vacuum and plasma, respectively, and the white diamond and red cross represent the launch position and cutoff location, respectively. These correspond to \textit{Scotty} beam-tracing runs for a probe beam with frequency $f=40$ GHz, and elevation and azimuthal launch angles $(\varphi_{elev},\ \varphi_{azim})=(0^\circ,\ 3^\circ)$, which achieves low mismatch at the cutoff $|\theta_{m,c}|\sim0^\circ$. Note that the probe beam is launched from and propagates only in the $Z=0$, XY-midplane cross-section, but that the viewing perspective is angled from the top.}
    \label{fig:flux_surfaces}
\end{figure}

In this paper, we use cylindrical coordinates $(R,\theta,Z)$, with $Z$ aligned with the long axis of the mirror and $R$ denoting the radial distance from this axis, see Figure~\ref{fig:flux_surfaces}. At times, we also use the associated Cartesian system, $(X,Y,Z)$. The high-temperature superconducting mirror coils are centered at the $Z=\pm 0.98~\mathrm{m}$ positions. \cite{Endrizzi:WHAM:2023}

The principal location of interest for this design is the central midplane, $Z=0$, where the magnetic field strength is minimum and the plasma cross-section is largest. This location provides direct access to the outer core, $0.7 \lesssim \rho \lesssim 0.9$, where flute-like density perturbations are expected to be observable. The fast-ion turning-point region is also of interest, since sloshing-ion pressure and energy can peak away from the midplane in neutral-beam-heated mirror plasmas. \cite{Endrizzi:WHAM:2023} In this work, however, we focus on the midplane port because it provides the most practical access for an initial DBS installation.

The midplane is also favorable from the perspective of mismatch. Since the equilibrium magnetic field there is predominantly axial, a beam launched from a midplane port can reach cutoff with its wavevector nearly perpendicular to the magnetic field. This geometry reduces mismatch attenuation and therefore maximizes the expected DBS signal. \cite{Hall-Chen:beam:2022, Hall-Chen:validation:2022} While small, there is also a suitable midplane port window available.

Note that in this work we use the angular convention $\varphi_{elev}$ for the elevation launch angle and $\varphi_{azim}$ for the azimuthal launch angle. Here, positive elevation launch angle $\varphi_{elev} > 0^\circ$ corresponds to the -Z-direction, and positive azimuthal launch angle $\varphi_{azim} > 0^\circ$ corresponds to the +Y-direction, see Figure~\ref{fig:flux_surfaces}.

\section{Measurement locations and wavenumbers} \label{sec:measurement}
We now determine the probe-beam frequencies and launch angles required to measure density fluctuations in WHAM. For DBS, the cutoff location sets the approximate measurement location, while the probe beam's wavevector at cutoff determines the measured fluctuation wavevector through the Bragg condition, $k_\perp \simeq 2 K_c$. Additionally, the probe beam's wavevector at cutoff should be perpendicular to the local magnetic field to maximize the backscattered signal. These requirements constrain the choice of antenna location, frequency range, and launch angles.

\begin{figure}
    \includegraphics[width=0.47\textwidth]{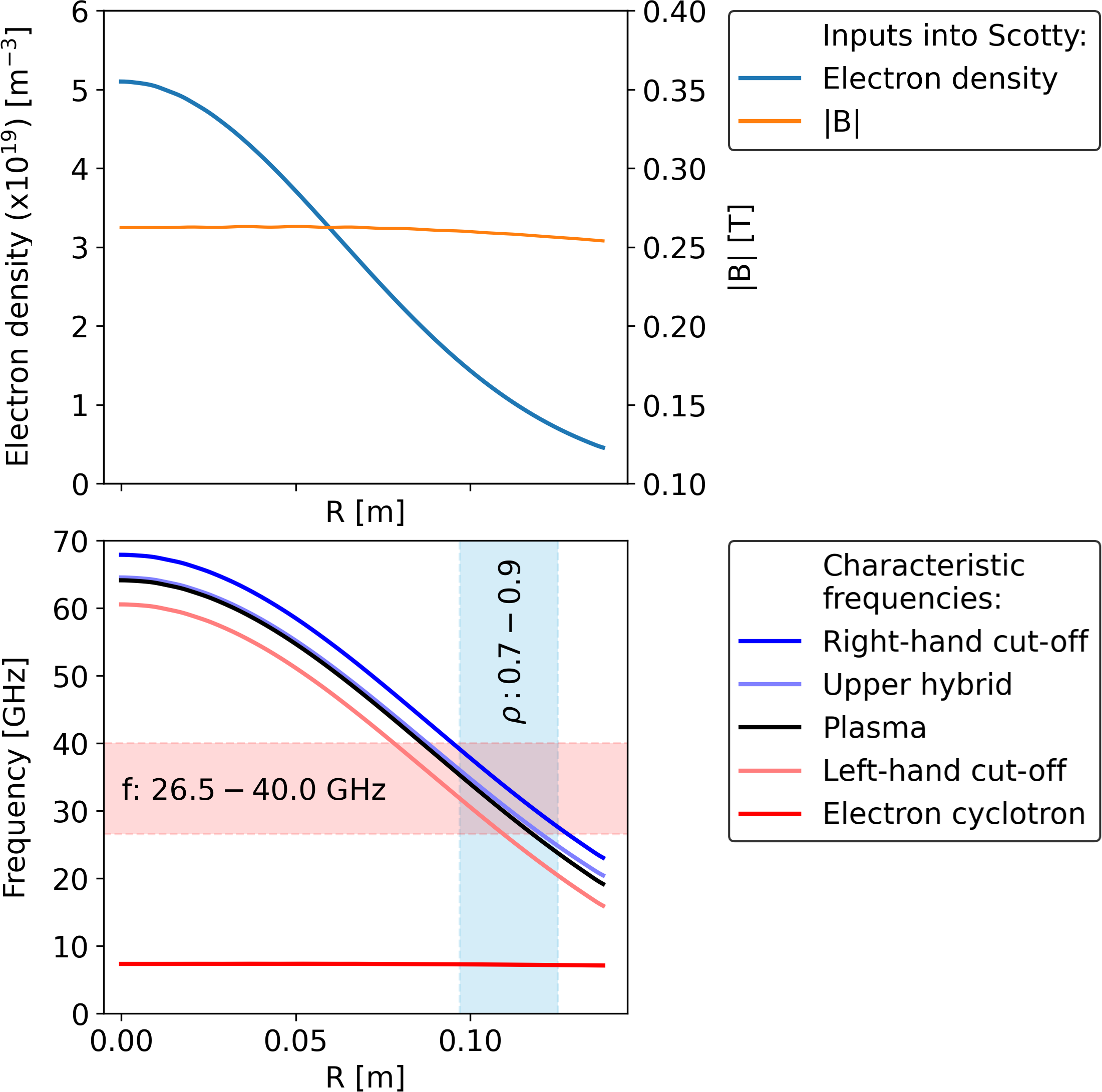}
    \caption{Electron density and magnetic field magnitude (top) and the characteristic frequencies (bottom), such as the plasma frequency, along the midplane, $Z=0$. The blue region (bottom) indicates the targeted measurement region, $0.7\lesssim\rho\lesssim0.9$, and the red region shows the Ka-band frequency range.}
    \label{fig:frequency_plots}
\end{figure}

Using cutoff frequencies along the midplane, $Z=0$, we estimated the range of frequencies required. For the representative WHAM equilibrium, the desired measurement region, $0.7\lesssim\rho_c\lesssim0.9$, corresponds to the X-mode cutoff frequencies in the Ka-band. Hence, we chose to use a tunable Ka-band source for the DBS system. Finally, absorption due to resonances, such as the electron cyclotron frequency and its second harmonic, is not expected. These characteristic frequencies are shown in Figure~\ref{fig:frequency_plots}.

\begin{figure}
    \includegraphics[width=0.47\textwidth]{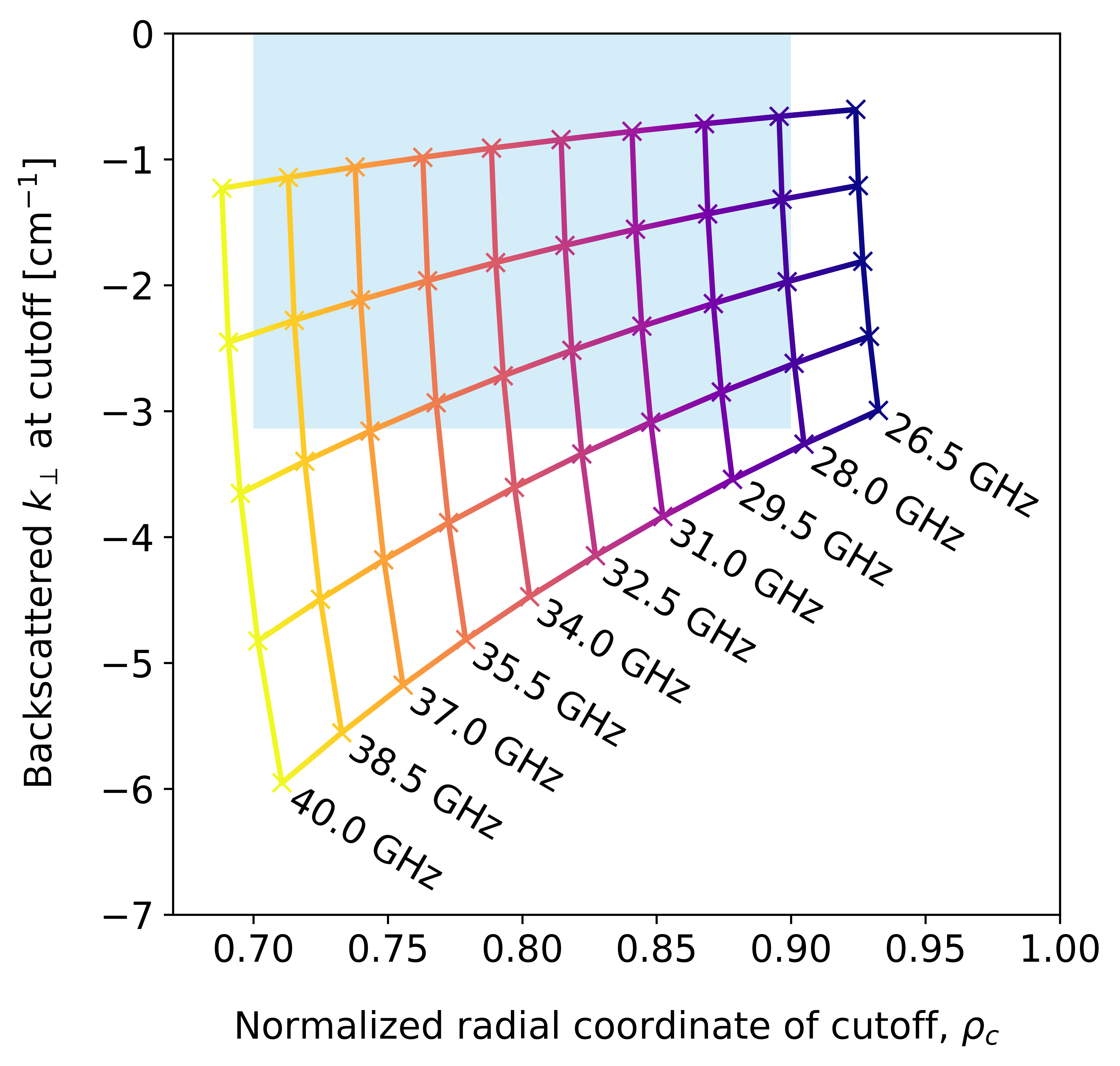}
    \caption{Backscattered wavenumber, $k_\perp$, as a function of the normalized cutoff location $\rho_c$ for azimuthal launch angles between $1^\circ$ and $5^\circ$. The blue region indicates the target measurement range, $1 \leq k_\perp \leq 3~\mathrm{cm}^{-1}$ and $0.7 \leq \rho_c \leq 0.9$. Beam-tracing simulations show that this range is accessed using probe frequencies between 28 and 38.5 GHz and azimuthal launch angles between $1^\circ$ and $3^\circ$.}
    \label{fig:kperp_vs_rho_c}
\end{figure}

Having narrowed down the frequency range, we determine the launch angles using \textit{Scotty} beam-tracing simulations. The elevation launch angle is fixed at $\varphi_{elev} = 0^\circ$ so that the beam remains in the midplane, where the magnetic field is predominantly axial, thereby naturally minimising mismatch. For each frequency, we varied the azimuthal launch angle, from $1^\circ$ to $5^\circ$. We found that the measurement location is predominantly determined by frequency and the measured wavenumber by the azimuthal launch angle, see Figure~\ref{fig:kperp_vs_rho_c}. 

As such, a fixed midplane launch between $1^\circ\leq\varphi_{azim}\leq3^\circ$ accesses the target edge region, $0.7\lesssim\rho_c\lesssim0.9$, where flute activity is expected to drive measurable shorter-wavelength density fluctuations. Since the preliminary quasioptical system has no steerable mirror due to space constraints, $\varphi_{azim}$ is set by rotating the external DBS assembly. The system is therefore fixed-angle during operation but reconfigurable during dedicated setup periods.

\section{Quasioptics and hardware} \label{sec:hardware}

\subsection{Quasioptical system}
The quasioptical system was designed to enable compatibility with the midplane port. The port window has a diameter of $9.5~\mathrm{cm}$, with a port thickness of $5.5~\mathrm{cm}$, making beam clearance the most challenging part of the DBS design. We therefore designed the quasioptics to minimize beam clipping at the port, requiring the beam to be focused near the center of the port window and to remain sufficiently smaller than the aperture throughout the port. While not further discussed in this paper, we have also used \textit{Scotty} to ensure that the beam properties are conducive to DBS performance.

We therefore consider a simple ex-vessel optical layout, consisting of a Ka-band horn followed by a single fixed focusing lens. Keeping the quasioptics outside the vacuum vessel enables fine-tuning during commissioning and avoids additional vacuum penetrations. The horn sets the initial Gaussian beam parameters, while the lens focuses the beam near the port window to maximize clearance through the narrow aperture. This quasioptical system avoids a steerable mirror and keeps the system compact, with the azimuthal launch angle set by rotating the DBS assembly.

\begin{figure}
    \includegraphics[width=0.47\textwidth]{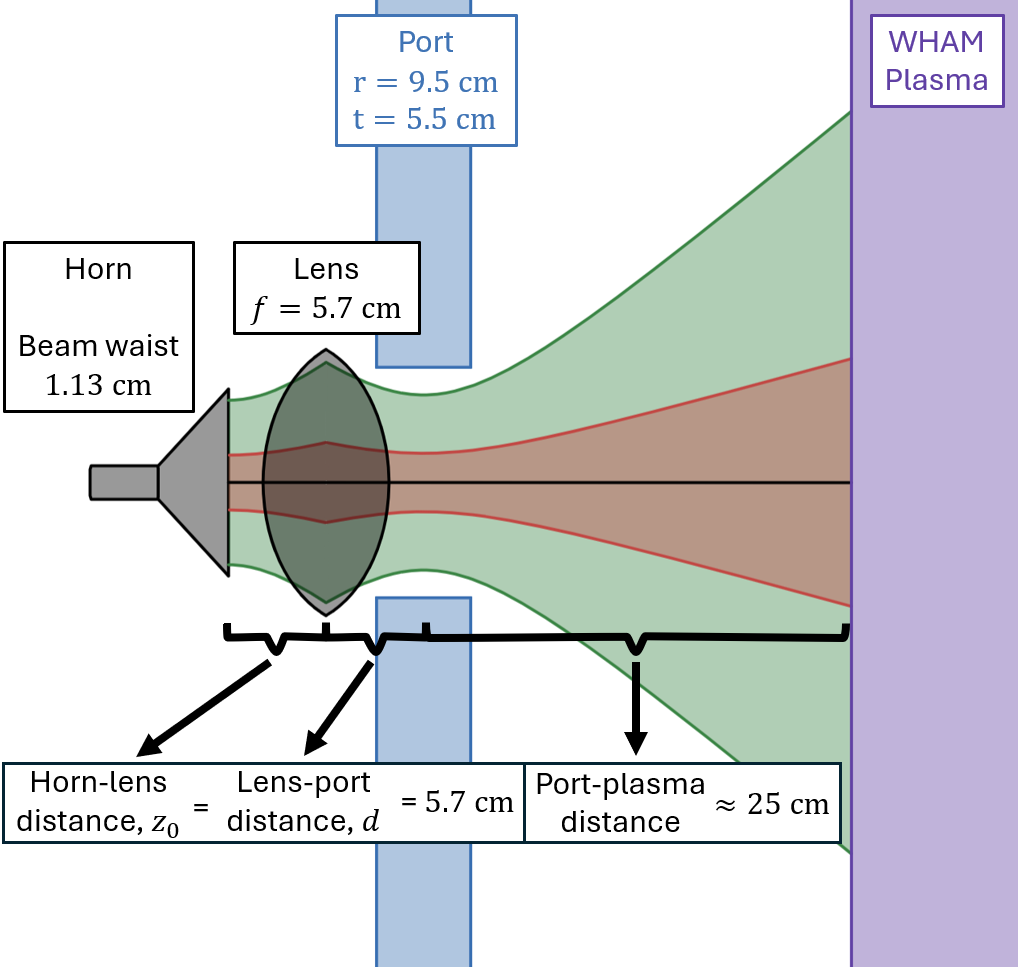}
    \caption{Schematic of the DBS quasioptics proposed for WHAM, visualised for a 40 GHz probe beam at elevation and azimuthal launch angles $0^\circ$, for illustration. The \textit{Scotty} beam-tracing code was used to model Gaussian beam propagation and explicitly calculate the beam width (red) and three-times-beam width (green). A parameter sweep of the DBS quasioptics configuration space led to the choice of the antenna-to-lens distance, $z_0$, lens-to-window distance, $d$, and the lens focal length $f$ to be 5.7 cm. This enables the Gaussian beam to fit completely within the lens (black) and port (blue), up to an azimuthal angle of $5^\circ$. Drawn to scale, except antenna size.}
    \label{fig:quasioptical_diagram}
\end{figure}

For the present design, an off-the-shelf Ka-band pyramidal horn and a biconvex ultra-high molecular weight polyethylene (UHMWPE) lens were found to be sufficient. The horn parameters used in the calculation are given in Table~\ref{tab:horn_lens_specs}. 
\begin{table}
    \begin{tabular}{|c|c|} \hline
        \multicolumn{2}{|c|}{\textbf{Horn specifications}} \\ \hline\hline
        Horn type      & Pyramidal horn         \\ \hline
        Beam radius at & For 26.50 GHz: 1.19 cm \\
        mouth of horn  & For 33.25 GHz: 1.17 cm \\
                       & For 40.00 GHz: 1.13 cm \\
                       & For other frequencies: \\
                       & linear interpolation performed \\ \hline
        Beam curvature & 0 cm$^{-1}$ \\
        at mouth of horn &           \\ \hline\hline
        \multicolumn{2}{|c|}{\textbf{Lens specifications}} \\ \hline\hline
        Lens type     & Biconvex                                   \\ \hline
        Lens material & Ultra-High Molecular Weight \\
                      & Polyethylene \cite{Xie:JTEXT-ECE:2020} (UHMWPE), $n=1.575$ \\ \hline
        Lens radius   & 5.5 cm \\ \hline
        Focal length  & 5.7 cm \\ \hline
    \end{tabular}
    \caption{Horn and lens properties of the quasioptical system. The horn specifications used are from an off-the-shelf horn, with a frequency range of 26.5--40 GHz and a gain of 20 dBi. In our work, we assume that the refractive index of the UHMWPE is independent of frequency.}
    \label{tab:horn_lens_specs}
\end{table}
In our work, we take the UHMWPE \cite{Xie:JTEXT-ECE:2020} lens to have a frequency-independent refractive index $n=1.575$, and thus yield lens dimensions which should be straightforward to fabricate. A parameter scan over the horn-lens distance, $z_0$, and lens focal length, $f$, shows that a compact solution can be obtained with $z_0=f=d=5.7~\mathrm{cm}$, where $d$ is the lens-window distance. This configuration keeps the beam within both the lens and port apertures, as shown in Figure~\ref{fig:quasioptical_diagram}. Final hardware testing will assess dielectric losses and the extent of reflections from the lens and window.

\subsection{Microwave hardware}
The proposed Ka-band microwave diagnostic is designed as a dual-purpose system: the primary mode is Doppler backscattering, while the same hardware can also be operated as a profile reflectometer. The latter capability is retained for flexibility, but is not analyzed in detail in this paper. Secondly, the microwave operating program must be changed, including the relationship between the transmission (Tx) and local oscillator (LO) frequencies, intermediate frequency (IF) bandwidth, and calibration procedure. In DBS mode, the system is configured to measure the power spectrum of the backscattered signal. In profile-reflectometry mode, it is configured to measure the phase delay of the reflected signal during a frequency sweep.

\begin{figure*}
    \includegraphics[width=0.98\textwidth]{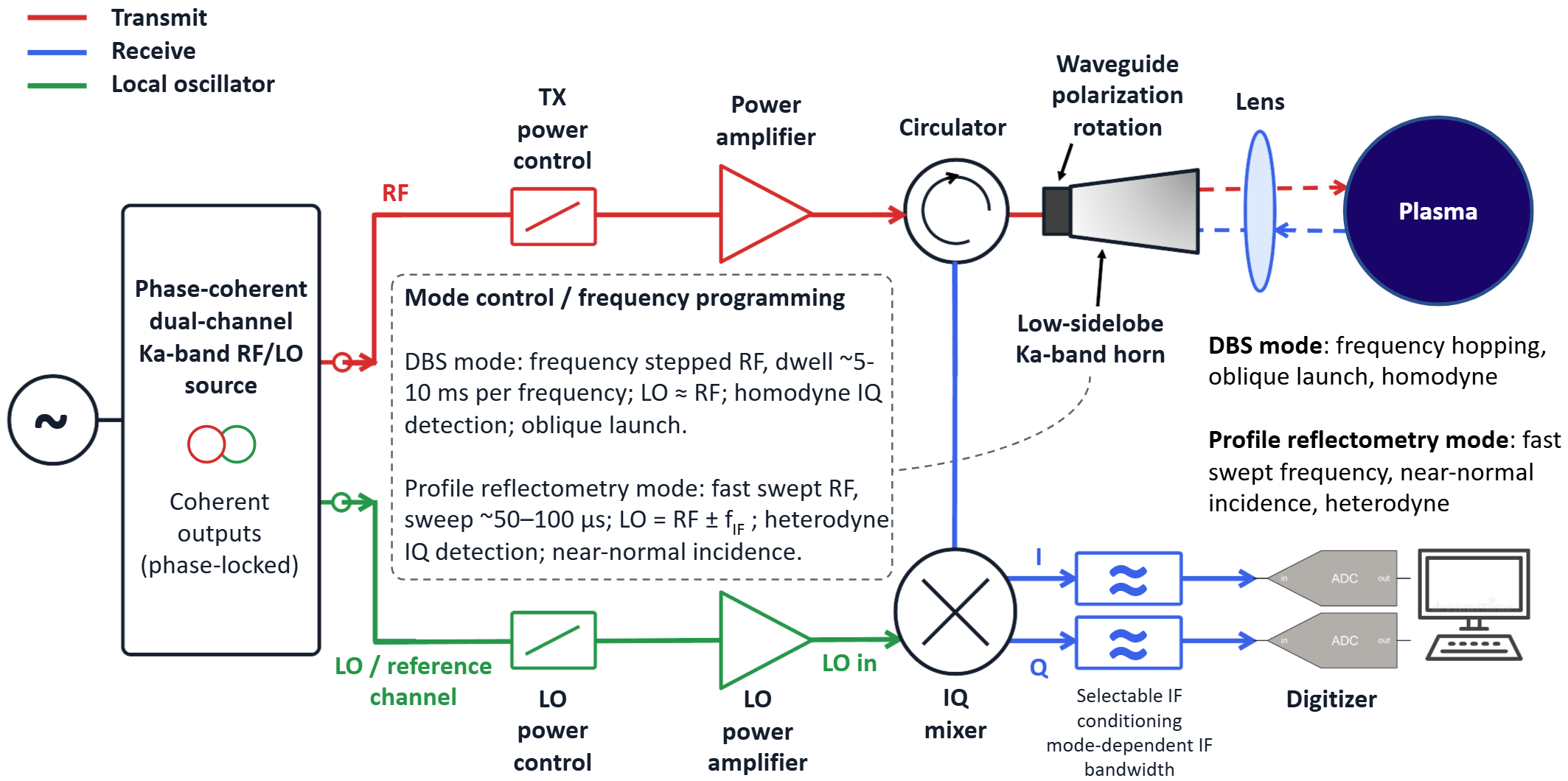}
    \caption{Block diagram of the dual-function Ka-band microwave diagnostic. Here, the transmit channel is referred to as TX, the intermediate frequency as IF, and local oscillator as LO. Note that these channels are phase-coherent.}
    \label{fig:block_diagram}
\end{figure*}

A phase-coherent dual-channel source --- one for TX and one for LO --- provides independent but phase-locked transmit and reference channels. The transmit signal is power controlled, amplified, and routed through a circulator to the Ka-band horn and quasioptics. The reflected or backscattered signal is collected through the same optical path, separated by the circulator, and mixed with the phase-coherent LO in an IQ mixer. The I and Q signals then pass through selectable IF conditioning before digitization. The shared hardware architecture is shown in Figure~\ref{fig:block_diagram}.
In DBS mode, the transmit channel operates in a stepped-frequency or frequency-hopping configuration. Each frequency is held for a dwell time of order several milliseconds, allowing the Doppler spectrum of the backscattered signal to be estimated. The LO is set close to the transmit frequency, enabling homodyne or near-homodyne IQ detection of the Doppler-shifted return signal.

In profile reflectometry mode, the same front-end hardware is operated in a swept-frequency configuration, like in typical reflectometry. \cite{Laviron:reflectometry:1996,Mazzucato:reflectometry:1998, Morales:reflectometry:2020, Rhodes:reflectometry:2024} The TX channel is swept across the selected Ka-band range, while the LO is swept coherently with a fixed frequency offset. This produces a heterodyne IF signal suitable for phase-delay and cutoff-position analysis. The intended sweep duration is approximately $50$--$100~\mu\mathrm{s}$, depending on the required radial resolution, plasma evolution time scale, and digitizer bandwidth. Note that swept-frequency profile-reflectometry operation requires sweep-linearity control, phase calibration, and appropriate operation-dependent IF and receiver settings. While our proposed architecture provides a route to profile-reflectometry operation, the detailed implementation, calibration, and operation-specific hardware optimization are beyond the scope of this paper.

Thus, the same Ka-band hardware platform can support both turbulence-sensitive DBS measurements and density-profile reflectometry. The mode-dependent parameters are the radio-frequency (RF) program, RF-LO frequency relationship, launch geometry, and IF bandwidth.

\section{Conclusions}
We designed a fixed-angle Ka-band dual-purpose microwave diagnostic for WHAM. The diagnostic is intended to operate primarily as a DBS system, providing spatially localized measurements of finite-$k_\perp$ density fluctuations and Doppler shifts associated with perpendicular flow. These measurements are aimed at assessing how fluctuations near the plasma boundary respond to applied biasing and sheared $E\times B$ rotation. The same hardware is intended to be compatible with profile reflectometer, providing density-profile information to support interpretation of the DBS measurements.

Using a representative experimentally-constrained WHAM equilibrium, we determined the measurement capabilities required of the diagnostic operating in DBS mode. With the \textit{Scotty} beam-tracing code, we calculated the X-mode cutoff locations, measured fluctuation wavenumbers, and mismatch angles for a range of probe frequencies and azimuthal launch angles. We found that a dual-channel tunable Ka-band system can access useful scattering locations from the deeper edge, $\rho_c\simeq0.7$, to near the last closed flux surface, $\rho_c\simeq0.9$, with measurable wavenumbers predicted to be $1\leq k_\perp\leq3~\mathrm{cm}^{-1}$. This region is expected to be relevant for density fluctuations driven or modified by flute activity. The selected midplane launch geometry also gives good magnetic-field matching at cutoff.

We also presented a preliminary quasioptical and hardware design compatible with the available WHAM midplane port. The proposed ex-vessel system uses a tunable Ka-band source, a horn, and a focusing lens to produce the Gaussian beam properties used in the beam-tracing calculations. Keeping the quasioptics outside the vacuum vessel allows mechanical fine-tuning, avoids additional vacuum penetrations, and enables timely initial installation. The design is frequency-tunable on the fly but fixed-angle during operation, with the azimuthal launch angle set by rotating the external assembly.

This diagnostic will contribute to local measurements of edge density fluctuations in WHAM. Such measurements will help assess how flute-active plasmas respond to applied biasing and sheared rotation, and will provide experimental constraints for validating stability and transport modelling in axisymmetric magnetic mirrors. With further refinements, the diagnostic may also have forward-looking applications to electron temperature gradient turbulence in tandem mirrors as well. \cite{Pratt:Tandem_confinement:2006, Frank:Tandem_performance:2025, Hillesheim:highk_measurement:2015}

\begin{acknowledgments}
This work was funded by Realta Fusion and carried out under a service agreement with the FEAT Centre, A*STAR, Singapore, and also partially funded by A*STAR: Strategic Tokamak Research for Industrial Deployment, and Energy (STRIDE) grant [H26-MSE152] and the FEAT-SRTT. The authors have no conflicts to disclose.

\textbf{Data availability statement}. The data that supports the findings of this study are available from the corresponding author upon reasonable request.

\textbf{AI declaration}. AI assistance was used to help draft and edit text, improve clarity and structure, and refine descriptions of the WHAM DBS design, including the equilibrium inputs, beam-tracing interpretation, quasioptics, hardware architecture, and conclusions. AI was not used to generate the scientific results, perform simulations, design the hardware, create figures, or determine the technical conclusions. All AI-assisted text was reviewed and edited by the authors before inclusion.
\end{acknowledgments}

\bibliography{aipsamp}

\end{document}